\documentclass{ws-ijmpa}
\usepackage[super,compress]{cite}
\usepackage{graphicx}

\def\beq{\begin{eqnarray}}
\def\eeq{\end{eqnarray}}

\def\al{\alpha}
\def\be{\beta}

\def\ga{\gamma}

\def\pa{\partial}

\def\si{\sigma}

\begin{document}
\markboth{Gon\c calves, Ribeiro, Pereira, and Dias Jr.}{The space-time torsion 
in the context of the Exact Foldy-Wouthuysen Transformation for a Dirac fermion}

%
\catchline{}{}{}{}{}
%

\title{The space-time torsion in the context of the Exact \\
 Foldy-Wouthuysen Transformation for a Dirac fermion
}

\author{Bruno Gon\c calves\footnote{
E-mail: bruno.goncalves@ifsudestemg.edu.br}
}
\address{Instituto Federal de Educa\c c\~ ao, Ci\^ encia e Tecnologia Sudeste de Minas Gerais \\
IF Sudeste MG, 36080-001, Juiz de Fora - MG, Brazil
}

\author{Baltazar J. Ribeiro\footnote{
E-mail: baltazarjonas@nepomuceno.cefetmg.br}
}
\address{Centro Federal de Educa\c c\~ ao Tecnol\' ogica de Minas Gerais \\
CEFET-MG, 37.250-000, Nepomuceno - MG, Brazil
}

\author{Dante D. Pereira \footnote{
E-mail: dante.pereira@cefet-rj.br}
}
\address{Centro Federal de Educa\c c\~ ao Tecnol\' ogica Celso Suckow da Fonseca\\
CEFET-RJ, 27.600-000, Valen\c ca - RJ, Brazil
}

\author{M\' ario M. Dias J\' unior\footnote{
E-mail: mardiasjr@yahoo.com.br}
}
\address{Instituto Federal de Educa\c c\~ ao, Ci\^ encia e Tecnologia Sudeste de Minas Gerais \\
IF Sudeste MG, 36080-001, Juiz de Fora - MG, Brazil
}

\maketitle

\begin{history}
\received{25 January 2016}
\accepted{12 April 2016}
Published  {05 May 2016}
\end{history}

\begin{abstract}
In this work we focus our attention in the inconsistency that appears when 
the Semi-Exact Foldy Wouthuysen transformation for the Dirac field interacting 
with space-time torsion field is performed. In order to solve this problem, we present 
a new involution operator that makes possible to perform the exact transformation when 
torsion field is present. Such operator has a structure, well known in the literature, 
composed of the product of an operator that acts in the matrices space and another one 
that acts in the function space. We also present the bound state of this theory and 
discuss the possible experimental analysis.

\end{abstract}

\ccode{PACS numbers: 03.65.-w; 11.15.Kc; 11.30.Er}



\section{Introduction}	

It is known that torsion fields arises when one takes into account 
the Gauge approach to gravity \cite{Utiyama,kibble}  
and this subject has received considerable attention of the scientific community. 
In Ref.~\refcite{shapiro1}, for example, it is possible to find a review about the
renormalization properties of quantum field theories in curved space-time in the presence
of CPT-Lorentz violating terms. Is also possible to find recent 
works (see Refs.~\refcite{baet1} and \refcite{CFMS} and references cited therein) that treat the 
possibility of CPT-Lorentz symmetry breaking in a more 
phenomenological point of view. Although there is not yet concise 
experimental evidences of torsion fields, it has been the aim of several 
recent studies (see Refs.~\citen{obukhov1,rydereshap,Hammond} and \refcite{Lammerzahl} 
and references cited therein). Spin-torsion discussion in the context of classical and quantum
effects are well described in Refs.~\refcite{ni} and \refcite{shapirorep}. 

As pointed out in Ref.~\refcite{obukhov1}, high energy level is more essential 
in the sense of getting torsion experimental manifestations. A concise review 
about Foldy-Wouthyusen Transformation (FWT) \cite{fw} and semi-classical limit 
for relativistic particles in strong external fields can be found in Ref.~\refcite{silenko1}.
Reference~\refcite{fw} also shows that FWT has succeeded in providing detailed information about
the nonrelativistic approximation. However, there is a considerable advantage in 
performing the Exact Foldy-Wouthyusen Transformation (EFWT) 
\cite{obukhov1,case,tiomo,silenko,eriksen,nikitin}. The reason is that, even EFWT is 
more complex\cite{eriksen,nikitin}, it presents some additional terms that can 
be missed if one uses FWT.
 
The magnitude of the coupling constant of the torsion field with the Dirac
spinor is very small\cite{shapirorep} and some features are specially 
related to a concise study of the nonrelativistic limit of the Dirac 
equation in the presence of an external torsion field. 
The case of the Dirac field interacting with many possible external fields associated 
with CPT-Lorentz violation was developed in the recent paper\cite{baltazar}, 
where the authors perform the EFWT together a review of the connection between 
CPT-Lorentz violating terms and phenomenology. 

Although the nonrelativistic limit was studied for the Dirac field 
interacting with the set of possible external fields (the torsion field could not be included in that set 
of external fields) associated with CPT-Lorentz violation\cite{baltazar}, 
there is not in the literature a concise study of the interaction of the Dirac field 
and the vectorial part of the torsion, in the context of the EFWT. In order 
to understand the reason for this we 
should mention that the possibility of performing the exact transformation depends on the fact that 
the commutation relation between the so called involution operator and the torsion 
field should be satisfied. One can check that for the torsion field, it is not. 
The first attempt in order to perform 
EFWT is known as Semi Exact Foldy-Wouthuysen Transformation (SEFWT)\cite{shapiro}. Such approach 
imposes some changes in the action of the theory, but it seems to work very 
well for several cases\cite{Bruno} and its results are in accordance with usual EFWT. 
However, when the torsion field is considered, SEFWT seems to fail\footnote{SEFWT consideration presents some 
not understandable physical impositions, for torsion case.}. 

In the present paper we consider the combined action of torsion and strong magnetic 
field on the massive spinor field and on the corresponding particle. 
In this case, the Hamiltonian does not admit the EFWT 
in the usual way. We discuss a method that enables one to perform EFWT and get physical results of this 
situation. We begin with the correct choice of involution operator. Such choice is quite natural in 
the sense that is in accordance with Refs.~\refcite{nikitin} and \refcite{violeta}. 
The basic idea is to work with an operator 
that acts only on the part of term that is not anti-commuting with the usual 
evolution operator. Taking into account that torsion breaks 
parity, a reasonable possibility would be to impose a new involution 
operator that has two contributions, the first one acts on matrices and another one on the functions. 
The second contribution of this operator (the new one) compensates 
the fact that the matrices commute with the involution operator. 

The method is used with the torsion field,  but it can be straightforwardly  
generalized to other terms. We emphasize that the method itself is the main results here, 
in the sense that it opens the window to the possibility of performing the EFWT for some 
cases that are until now, not contemplated by the literature and extract from its physical information. 
Experimental perspectives are also analyzed. 

The paper is organized as follows. In section \ref{sefwt} we present a brief 
review about SEFWT and discuss some restrictions of this model when torsion field 
is considered. In section \ref{opinovo}, the new proposal to perform EFWT for the Dirac field 
interacting with space-time torsion is presented. Sections \ref{eqofm} and \ref{boundstate}  
are devoted to the equations of motion and the study of the bound state of the theory, respectively.  
In section \ref{conc} we draw our conclusions. Throughout 
the paper we use Greek letters for the indexes which
run from 0 to 3. Latin indexes are used for the space coordinates and run from 1 to 3.


\section{Semi Exact Foldy-Wouthuysen Transformation}
\label{sefwt}

We present in this section a brief review about SEFWT\cite{shapiro,Bruno}. 
Consider the spin-$1/2$ particle in an external 
torsion and electromagnetic fields. We are going to consider 
the magnetic and torsion fields which can only vary with time, 
but do not depend on the space coordinates. The Hamiltonian we shall 
deal with is written as follows 
\begin{equation}
{\cal{H}}=c\overrightarrow{\alpha}\cdot\overrightarrow{p}-
e\overrightarrow{\alpha}\cdot\overrightarrow{A}-
\eta_1\overrightarrow{\alpha}\cdot\overrightarrow{S}\gamma_5+e\Phi+
\eta_1\gamma_{5} S_0+mc^2\beta \,.
\label{ham1}
\end{equation}

Here we used notations $A_\mu=(\Phi\,,\,\overrightarrow{A})$,
$S_\mu=(S_0\,,\,\overrightarrow{S})$. In case of constant magnetic 
field, one can set $\Phi=0$. We adopt notations as described in Ref.~\refcite{BD} 
for Dirac Matrices and also denote the $\ga^0$ Dirac matrix as $\be$. 

Only those theories where the 
Hamiltonian obey the following relation, enable one to perform the EFWT\cite{case,eriksen,nikitin,diraceq}.

\vspace{-2mm}
\begin{equation}
J{\cal{H}}+{\cal{H}}J = 0 \,,\label{involution}
\end{equation}

where $J=i\ga_{5}\be$. The quantity $J$ is the so called involution operator, which is Hermitian and unitary. 
It is also known that $J\be+\be J=0$.

Direct inspection show that the term $\eta_1\overrightarrow{\al}\cdot\overrightarrow{S}\ga_5$ 
is the only one in the Hamiltonian (\ref{ham1}) that does not satisfy the condition (\ref{involution}).
From this point of view, a natural conclusion is that would not be possible to perform EFWT when one 
take into account the torsion field in the Hamiltonian of the theory. However, 
there is a possible consideration that modifies this scenario, 
in some sense (see eg. Ref.~\refcite{shapiro} and references cited therein). 
Let us make an {\it ad hoc} modification. According to this modification, the term commented above  
should be multiplied by the $\be$-matrix. Observe that such modification satisfies 
the condition (\ref{involution}) and now the EFWT is perfectly possible\footnote{In the linear order in the
torsion field, an extra $\be$ has no effect.}. After all, the Hamiltonian we are going 
to deal with has the form
\beq
{\cal{H}}\,=\,c\overrightarrow{\al}\cdot\overrightarrow{p}
- e\overrightarrow{\al}\cdot\overrightarrow{A}
- \,\eta_1\overrightarrow{\al}\cdot\overrightarrow{S}\ga_5\be
\,+ \eta_1\ga_{5} S_0+mc^2\be\,.
\label{H}
\eeq

According to the standard EFWT prescription\cite{eriksen,baltazar}, 
the next step is to obtain $H^2$. Direct calculations give the result
\beq
{\cal{H}}^2 &=& (c\overrightarrow{p}-
e\overrightarrow{A}-
\eta_1\overrightarrow{\Sigma}S_0)^2+
m^2c^4+2\eta_1mc^2\overrightarrow{\Sigma}\cdot\overrightarrow{S}
\nonumber
\\
&-& (\eta_1)^2(\overrightarrow{S})^2-
\hbar ce\overrightarrow{\Sigma}\cdot\overrightarrow{B}-2(\eta_1)^2(S_0)^2
+i\eta_1\gamma_5\beta\overrightarrow{\Sigma}\cdot
\big[\overrightarrow{S}\times
(c\overrightarrow{p}-e\overrightarrow{A})\big]
\,\,.
\label{h2comp}
\eeq

Observe that the last term in this equation transforms (under parity) 
in a different way compared to the other terms 
in the Hamiltonian. However, there is no reasonable physical arguments that 
enable us to suppose that 
$\overrightarrow{\Sigma}\cdot [\overrightarrow{S}\times 
(c\overrightarrow{p}-e\overrightarrow{A})]=0$. From this point, the next step 
is to perform the exact transformation, We shall not to describe this procedure 
in details here (standard procedure is described in Refs.~\refcite{shapiro} and 
\refcite{diraceq}. The transformed Hamiltonian is written as follows
\beq
{\cal{H}}^{tr} &=& 
\beta mc^2+\frac{\beta}{2mc^2}(c\overrightarrow{p}
- e\overrightarrow{A}- \eta_1\overrightarrow{\Sigma}S_0)^2
+ \beta\eta_1\overrightarrow{\Sigma}\cdot\overrightarrow{S}
\nonumber
\\
&-& \beta\frac{\hbar e}{2mc}\overrightarrow{\Sigma}\cdot\overrightarrow{B}
-\beta\frac{(\eta_1)^2}{mc^2}(S_0)^2
+i\beta\eta_1\gamma_5\beta\overrightarrow{\Sigma}\cdot
\big[\overrightarrow{S}\times
(c\overrightarrow{p}-e\overrightarrow{A})\big]
\,.
\label{oldham}
\eeq

Now and so on we denote the terms with ''tr'' index as the transformed ones
and such terms belong to the final transformed Hamiltonian. 
Taking into account the two components spinor 
\begin{equation}
\psi  = 
\left( \begin{array}{c}
\varphi  \\
\chi    \\
       \end{array}
\right)
e^{\frac{-imc^2t}{\hbar}}
\label{expo}
\end{equation}

and writing the Dirac equation in the Schr\"{o}dinger form $i\hbar\pa_t\psi={\cal{H}}\psi$, 
the Hamiltonian for $\varphi$ is written in the following way

\begin{equation}
{\cal{H}}^{tr}_\varphi=\frac{1}{2m}(\overrightarrow{\Pi})^2+B_0+
\overrightarrow{\sigma}\cdot\overrightarrow{Q}\,,
\label{ham99}
\end{equation}

where

\begin{eqnarray}
\overrightarrow{\Pi}&=&\overrightarrow{p}-
\frac{e}{c}\overrightarrow{A}-
\frac{\eta_1}{c} S_0 \overrightarrow{\sigma}
\quad,\quad
B_0=-\frac{(\eta_1)}{mc^2}^2(S_0)^2\nonumber\\\nonumber
\overrightarrow{Q}&=&\eta_1\overrightarrow{S}
-\frac{\hbar e}{2mc}\overrightarrow{B}+\frac{\eta_1}{mc}
\overrightarrow{S}\times(\overrightarrow{p}-\frac{e}{c}\overrightarrow{A})\,.
\label{ham100}
\end{eqnarray}

The canonical quantization of (\ref{ham99}) lead us to (quasi) classical equations of motion

\begin{eqnarray}
\frac{dx_i}{dt}=\frac{1}{m}\Big(p_i-\frac{e}{c}A_i-\frac{\eta_1}{c}\si_i S_0\Big)
+\frac{\eta_1}{mc}\Big[\overrightarrow{\si}\times\overrightarrow{S}\Big]_i=v_i \label{eqmov1}\\
\frac{dp_i}{dt}=\frac{1}{m}\Big(p^j-\frac{e}{c}A^j-\frac{\eta_1}{c}\si^j S_0\Big)
\frac{e}{c}\frac{\pa A_j}{\pa x^i}+\frac{\eta_1}{mc}\Big[\overrightarrow{\si}\times\overrightarrow{S}\Big]^j
\,\frac{e}{c}\frac{\pa A_j}{\pa x^i}
\label{eqmov2}
\end{eqnarray}
\begin{equation}
\frac{d\si_i}{dt}=\Big[\overrightarrow{R}\times\overrightarrow{\si}\Big]_i
\hspace{0.2cm}\mbox{,}\hspace{0.2cm}
{R_j}=\frac{2\eta_1}{\hbar}\Big[{S_j}-\frac{1}{c}{v_j}S_0
+\Big({S}\times\frac{\overrightarrow{v}}{c}\Big)_j+\frac{2\eta_1}{\hbar}S_0\Big(\overrightarrow{S}
\times\overrightarrow{\si}\Big)_j\Big]+\frac{e}{mc}{B_j}\,.
\label{eqmov3}
\end{equation}

Combining these last equation, the Lorentz force is written as
\begin{equation}
m\frac{dv_i}{dt}=-\frac{e}{c}\frac{\pa A_i}{\pa t}+\frac{e}{c}\Big[\overrightarrow{v}
\times\overrightarrow{B}\Big]_i
-\frac{\eta_1}{c}\si_i\frac{\pa S_0}{\pa t}-\frac{\eta_1}{c}
\frac{\pa (\overrightarrow{S}\times\overrightarrow{\si})_i}{\pa t}
\label{eqmov4}
\end{equation}

Based on what was explained above, one can tend to suppose that the SEFWT 
approach is not consistent. On the other hand, the SEFWT is performed in Ref.~\refcite{Bruno} 
for several cases and the results are in accordance with usual EFWT. 
Although the approach itself seems not to have inconsistencies, it fails, in the practical sense 
for the case studied here. In order to get a better perspective about this situation, 
we present in the next section a new proposal to perform the EFWT transformation 
for the space time torsion case. 


\section{Exact Foldy-Wouthuysen Transformation, the new proposal}
\label{opinovo}

We present here an approach that enables one to work with the usual EFWT 
for the torsion field. The main idea is to consider a more general involution operator form
rather than the one used in the previous section. We shall consider the more general 
involution operator structure\cite{nikitin,violeta}
\vspace{-1mm}
\begin{equation}
J=M \times \hat{F}\,,
\label{choo}
\end{equation}

where $M$ and $\hat{F}$ are operators that act on the matrices and functions 
(external fields in the action for example) space respectively. 
With this assumption the general form of the Hamiltonian (\ref{ham1}) is not changed. The 
involution operation we shall deal with has the following explicit form 
\begin{equation}
J=i\ga^5\be \,\hat{P} \,\hat{T}
\label{novop}
\end{equation}

where $\hat{T}$ is time reverse operator and $\hat{P}$ the parity operator.

One can find in the introduction of Ref.~\refcite{kostWIEN} a list 
of references to CPT theorem. It is important to remember some basic relations for the parity
reflection $\hat{P}$ and time reversal $\hat{T}$ that are important for us in this work 
for quadri-vectors. The important thing here is to take into account how the 
vectors and pseudo-vectors respond to the action of these operators. The main 
point is that under T-transformation only the time component of the four 
vector changes sign and for the P-transformation the vector part is affected. For a pseudo-vector, 
like $S_\mu$, the situation is that if $\;x_i^\prime \,\rightarrow\, -x $ (parity), the vector part 
changes sign and if $\;t^\prime \,\rightarrow\, -t $, the $S_0$ part changes sign. As it should be, 
since we don't have C-symmetry breaking terms of this Hamiltonian, for this 
case the transformation PT will give the covariance of the Hamiltonian. 

Therefore, what we are proposing here that can be considered a new approach is a method to 
find the correct form of the involution operator that allows the EFWT method to be applied 
in some cases it would not be possible. Here, the involution operator (\ref{novop}), that 
has the same form, for example, in Ref.~\refcite {nikitin}, does not the restrict 
the form of the external analyzed field, as it was done for the electromagnetic potential 
vector on the cited work. The idea here is applied only for possible CPT/Lorentz symmetry 
breaking terms. One should know which kind of symmetry the studied term breaks, before the 
calculations (from the literature). In our case we have parity, for torsion, as an example. 
Then the next step is to propose a form for the operator $\hat{F}$ in ($\ref{choo}$) that is 
$\hat{P}\, \hat{T}$, in our case.   

Observe now that the commutation relation (\ref{involution}) is obeyed, when one 
take into account relations (\ref{novop}) and the Hamiltonian of the system (\ref{ham1}). 
For this reason, EFWT is completely possible to be performed. It is worth noting that in a 
general case, if one wants to perform the exact transformation or any 
external term, what need to be done is to find the explicit form for operator $\hat{F}$. 
It must be done to show the consistence of the method. 

At this point we are able to perform the EFWT. The procedure we use is the standard one, 
which is well described in Refs.~\refcite{baltazar,shapiro} and \refcite{diraceq}. The 
transformed Hamiltonian, for the Dirac spinor is written as follows 
\beq
{\cal{H}}^{tr} &=& 
\beta mc^2+\frac{\beta}{2mc^2}(c\overrightarrow{p}
- e\overrightarrow{A}- \eta_1\overrightarrow{\Sigma}S_0
- \eta_1 \gamma_5 \overrightarrow{S})^2
+ \beta\eta_1\overrightarrow{\Sigma}\cdot\overrightarrow{S}
\cr
&-& \beta\frac{\hbar e}{2mc}\overrightarrow{\Sigma}\cdot\overrightarrow{B}
-\beta\frac{(\eta_1)^2}{mc^2}(S_0)^2
+\beta\frac{(\eta_1)^2}{2mc^2}(\overrightarrow{S})^2\,.
\label{novaham}
\eeq

We remark that this last equation is completely free of breaking parity terms. 
Nevertheless, a comparison between the equations (\ref{oldham}) and (\ref{novaham}) 
shows that the Hamiltonian described by (\ref{novaham}) presents a torsion vector 
contribution in the kinetic part.


\section{Equations of motion}
\label{eqofm}

We perform in this section the calculations of equations of motion. 
Let us begin by  taking into account the two components spinor, described by (\ref{expo}). 
As explained in section (\ref{sefwt}), the next step is to write the Dirac equation in the 
Schr\"{o}dinger form $i\hbar\,\pa_t\psi={\cal{H}}\psi$. Straightforward calculations 
enable one to write the Hamiltonian for $\varphi$ as 
\begin{equation}
{\cal{H}}^{tr}_\varphi=\frac{1}{2m}(\overrightarrow{\Pi})^2+B_0+
\overrightarrow{\sigma}\cdot\overrightarrow{Q}\,,
\label{hamtr}
\end{equation}

where
\begin{equation}
\overrightarrow{\Pi}=\overrightarrow{p}-
\frac{e}{c}\overrightarrow{A}-
\frac{\eta_1}{c} S_0 \overrightarrow{\sigma}-
\frac{\eta_1}{c} \sigma_5 \overrightarrow{S}
\hspace{0.7cm}\mbox{,}\hspace{0.7cm}
B_0=-\frac{(\eta_1)}{mc^2}^2(S_0)^2+
\frac{(\eta_1)}{2mc^2}^2(\overrightarrow{S})^2
\, ,\nonumber\\
\end{equation}
\begin{equation}
\overrightarrow{Q}=\eta_1\overrightarrow{S}
-\frac{\hbar e}{2mc}\overrightarrow{B}\,,
\label{ham5}
\end{equation}

where $\sigma_5 \,=\,({\scriptstyle 1/6})\,\varepsilon^{ijk} \, \sigma_i \, \sigma_j \, \sigma_k$, 
see Ref.~\refcite{diraceq}.The expressions above are not exactly the same as derived in 
Refs.~\refcite{rydereshap} and \refcite{bagrov} through the usual perturbative FWT. The basic difference
is the term $\si_5 \overrightarrow{S}$. The appearance of this new term is based on advantage of using 
EFWT instead of FWT\footnote{Using the Exact transformation, the risk of missing some important terms is lower.}.

It is important to note that the presence of terms of the kind $\overrightarrow{S} \, \cdot \, \overrightarrow{B}$ 
in the transformed Hamiltonian (\ref{hamtr}) is related to the possibility of considering experimental tests of 
torsion field using magnetic resonance, as it was explained in \cite{bagrov}. Nevertheless, a straightforward 
comparison between equations (\ref{ham99}) and (\ref{hamtr}) shows two differences between the SEFWT approach 
and the method presented here. The first one represents a new contribution in the kinetic part of (\ref{hamtr})
represented by a term\footnote{Observe that such term is a new one with relation to FWT\cite{rydereshap} and 
SEFWT\cite{shapiro}.} of the kind $\si_5 \overrightarrow{S}$. The second one is the absence, in the Hamiltonian 
(\ref{hamtr}), of a breaking parity term. 

In order to quantize the Hamiltonian (\ref{ham5}) and to write semi-classical 
equations of motion (After the calculus we make $\hbar \rightarrow 0$ in 
the same procedure adopted in Ref.~\refcite{bush}). Let us consider the following relations
\begin{equation}
i\hbar \frac{d \hat{x}_i}{dt}=[\hat{x}_i,{\cal H}],\hspace{1cm}
i\hbar \frac{d \hat{p}_i}{dt}=[\hat{p}_i,{\cal H}]\label{quant}\hspace{1cm}
\mbox{and}\hspace{1cm}i\hbar\frac{d\,\hat{\si}_i}{dt}=[\hat{\si}_i,{\cal H}]\,.
\end{equation}

So we get
\begin{eqnarray}
\frac{d\hat{x}_i}{dt}&=&\frac{1}{m}\Big({p_i}-\frac{e}{c}{A_i}
-\frac{\eta_1}{c}{\si_i} S_0-\frac{\eta_1}{c} \si_5 {S_i} \Big)\,=\,v_i\label{mov1}\\
\frac{d\hat{p}_i}{dt}&=&\frac{\pi^j}{mc}\Bigg(e\frac{\pa A_j}{\pa x^i}
+\eta_1 S_0\frac{\pa \si_j}{\pa x^i}+
\eta_1 S_j\frac{\pa \si_5}{\pa x^i} \Bigg)\label{mov2}\\
\frac{d\hat{\si}_i}{dt}&=&\Big[\overrightarrow{R}\times\overrightarrow{\si}\Big]_i\,,
\label{mov3}
\end{eqnarray}

where 

\begin{eqnarray}
R_j&=&2\frac{\eta_1}{\hbar}\Big[S_j-\frac{1}{c}v_j S_0\Big]
-\frac{e}{mc}B_j\,,
\end{eqnarray}

and $\si_5$ is the $\ga_5$ representation for the bi-spinor. 

Therefore,
\begin{equation}
m\frac{dv_{i}}{dt}=\Big[\overrightarrow{v}\times\overrightarrow{C}\Big]_i+\frac{d}{dt}\big({u_i}\big)\,,
\label{mov4}
\end{equation}

where

\begin{equation}
C_k=-\frac{e}{c}B_k-\frac{\eta_1}{c}
\varepsilon_{klm}{\frac{\pa}{\pa x_l}}{\big(S_0\si^{m} + \si_{5}S^m\big)}
\hspace{0.8cm}\mbox{and}\hspace{0.8cm}
u_i=-\frac{e}{c}A_{i}-\frac{\eta_{1}}{c}\big(S_{0} \si_{i} - \si_{5}S_i\big)\,.
\end{equation}

The equation presented above represents the corrections for the classical Lorentz
force acting on the Dirac particle. If one considers a trajectory described by this fermion, it is possible 
to observe that the terms with $S_\mu$ could offer corrections for the path of the particle. 
These results are in accordance with the known equations of motion presented on Ref.~\refcite{rydereshap}. 

Comparing the results for the equations of motion, that means, in this case to compare the exact approach with the 
semi-exact one, it is possible to see some differences. Looking one by one, we can note that the terms with $S_i$ 
have different algebraic construction in (\ref{eqmov1}) and (\ref{mov1}). But in both equations they have the same 
physical meaning since it is mixed with the spinor matrices in first order (that is what matters for our 
phenomenological approach). Analogous considerations can be performed for equations (\ref{eqmov2}) and (\ref{mov2}), 
in which the unique difference is in the terms with $\sigma_i$ and $S_i$. Finally, the equations (\ref{eqmov3}) and 
(\ref{mov3}) have no difference at all if we look for them carefully. The term of second order in torsion in 
(\ref{eqmov3}) was considered neglectable in (\ref{mov3}). The term with the vector product between $S_i$ and $v_i$ 
are zero because if we simply substitute $v_i$ from (\ref{mov1}) into this term,  we can see that the term with $A_i$ 
has the factor $v/c^2$ (we are dealing with the nonrelativistic limit of the theory) and the others contribute only 
for the second order in torsion field. The term with the pure spatial part of momentum $p_i$ will produce no 
physical difference when multiplied for the terms with torsion since each of these terms have a derivative of 
spin matrices with respect to the coordinates (It does not contribute for the trajectory of the particle, as it 
can be seen on equations (\ref{eqmov4}) and (\ref{mov4})).  

Another point that must be empathized is the necessity to extract from the exact transformed Hamiltonian 
the bound state of the theory, in order to propose possible experimental tests. As we know, the bound state would 
give us the possibility to use the powerful method presented in the series 
of papers\cite{coll+kost, kostelecky21, costelechyprd, costelechy2, costelchy1} to find 
another possible experimental text for the torsion field using this theory. In the next section we present 
some comments and calculations about this relevant subject.


\section{Bound state considerations}
\label{boundstate}

In this section we present brief considerations about 
the bound state of the Dirac Field interacting with 
space-time torsion. The perspective of CPT-Lorentz violation tests has considerable 
advantages in the context of Quantum Electrodynamics systems. There is, in fact, 
a set of examples related to atomic physics experiments (See Refs.~\refcite{blumblumpapers} and \refcite{lane}, 
and references cited therein). In this sense, it is completely relevant the calculation of the bound state of 
the theory\cite{costelchy1} we are considering here.

Let us consider the Lorentz violating potential $V$ given by the following relation\cite{lane}

\vspace{-5mm}
\begin{equation}
V=-\tilde{b}_j \si^{j}\,,
\end{equation}

where $\si$ is the spin matrices. The Lorentz potential comes from the equation 
(\ref{hamtr}) and considering such equation one can write the following bound state 

\begin{equation}
\tilde{b}_j = b_j - \eta_1 S_j+\frac{\hbar e}{2mc}B_j\,. 
\label{bound}
\end{equation}

In the last equation we observe the torsion contribution to the bound state. Such contribution is 
completely new and was not contemplated in the bound state associated to the EFWT for
a Dirac theory related to the 80 CPT-Lorentz violating terms\footnote{The reason is that, in Ref.~\refcite{baltazar}
the criteria to perform EFWT is anti-commutation relation between the Hamiltonian and $i\ga^5\be$.}\cite{baltazar}. 
However, although the possibility of indications of possible atomic experiments\cite{costelechy2,costelchy1} is 
related to the bound state (\ref{bound}), the magnitude of torsion field is irrelevant when compared, for example, 
with the magnitude of the magnetic field. For this reason, a concise proposal about experimental measurements of 
torsion field is not straightforward. 


\section{Discussion and conclusion}
\label{conc}

The nonrelativistic limit has been already studied for the Dirac field 
interacting with a set of external fields (except for the torsion field) 
in the context of CPT-Lorentz violation\cite{baltazar}. However, 
torsion case does not admit the usual exact transformation and the semi-exact 
transformation also seems to fail in such case. In this paper, 
we have derived a special technique that enables one 
to perform EFWT for the Dirac spinor field in the combined background of torsion
and constant uniform magnetic fields. The Hamiltonian corresponding to the 
nonrelativistic limit was presented (\ref{hamtr}) and as one can check, it is 
completely free of braking parity terms.

We also have derived the equations of motion (\ref{mov1}), (\ref{mov2}) and (\ref{mov3}) 
for the situation described above and the Lorentz force corrected by the presence of torsion 
field (\ref{mov4}) was presented, together the discussion of possible experimental manifestations. 
We have calculated bound state of the Dirac field interacting with 
space-time torsion represented by the equation (\ref{bound}). However, due to the weakness 
of the torsion field, there is no a final conclusion that point out to the perspective 
about measuring the torsion field, using this technique. Notwithstanding, the main result here is the method itself, 
since it can be straightforwardly generalized in order to perform the EFWT for several 
cases until now not contemplated in the literature. 

It is remarkable to say that the method presented in this work gives the possibility of performing 
the exact transformation for external fields not contemplated in Ref.~\refcite{baltazar}. 
In general, for each external field in the Hamiltonian (when performing EFWT is not possible) 
of the theory, there should be a particular special involution operator of the kind described in 
the equation (\ref{choo}). In this work we have considered torsion field.
However, the search for such operators is a hard task and the study of a more general involution operator 
that contemplates all the possible external fields mentioned above should be in development, in a near future.


\section*{Acknowledgments}

BG and MJ are grateful to Funda\c c\~ ao Nacional de Desenvolvimento 
da Educa\c c\~ ao (FNDE) for financial support.



\end{document}